# TOWARDS MEASURING AND SCORING SPEAKER DIARIZATION FAIRNESS

*Yannis Tevissen[1,2], Jérôme Boudy[1], Gérard Chollet[1], Frédéric Petitpont[2]*

[1] SAMOVAR, Telecom SudParis, Institut Polytechnique de Paris, France
[2] Newsbridge, France

## ABSTRACT

Speaker diarization, or the task of finding "who spoke and when", is now used in almost every speech processing application. Nevertheless, its fairness has not yet been evaluated because there was no protocol to study its biases one by one. In this paper we propose a protocol and a scoring method designed to evaluate speaker diarization fairness. This protocol is applied on a large dataset of spoken utterances and report the performances of speaker diarization depending on the gender, the age, the accent of the speaker and the length of the spoken sentence. Some biases induced by the gender, or the accent of the speaker were identified when we applied a state-of-the-art speaker diarization method.

***Index Terms***— speaker diarization, fairness, ethical AI, evaluation protocol

## 1. INTRODUCTION

Speaker diarization is the task of determining "who spoke when" in an audio or video stream. It is widely used in speech recognition systems that need to analyze multiple speaker conversations. Several approaches and architectures have been developed [1] in the past decade and the field has known a renewed enthusiasm with the rise of deep learning architectures [2].

Today's speaker diarization is used for a wide variety of tasks and in a wide variety of scenarios, but the overall principles often remain the same. A standard speaker diarization algorithm is made of a voice activity detection and segmentation module, followed by a speech embedding extractor and finally a clustering phase.

Recent research aims at improving speaker diarization robustness in the most complex cases. Major advances have been done in situations of overlapping speech [3]–[5], dinner party diarization [6] and unlimited number of speaker handling [7].

## 2. FAIRNESS IN SPEECH PROCESSING

While speaker diarization, speech recognition and more generally machine learning applications are entering our daily lives, the fairness of these systems is yet to be proven.

Fairness has become a major topic for AI research, starting with computer vision [8]. Major applications used by millions have been proven to be biased especially on ethnical, gender, and age characteristics [9].

Some research has recently been performed on speech recognition fairness [10], [11] and highlighted notable biases in the most used algorithm. For instance, word error rates of some commercial ASR systems were demonstrated to be significatively higher for black speakers than for white speakers.

In the meantime, more and more datasets are made available with the proper labels to study these biases [12], [13].

Although recent research on speaker diarization focuses on its robustness in the most complex scenarios, there have been no proper study of the fairness of speaker diarization state-of-the-art methods.

## 3. PROPOSED PROTOCOL

### 3.1. Tracking biases in Speaker Diarization

It is particularly complex to study the biases of a speaker diarization system. Indeed, as the inputs of such algorithms are conversations, it is very hard to replicate the same conditions between two recordings. Every conversation is unique, and many factors must be considered among which the speaker turns, the length of the sentences and the similarity between the voices of the two (or more) speakers. As far as we know there are no datasets to study the fairness in speaker diarization in a multi-speaker scenario. The most used diarization datasets [14]–[16] contain natural or scripted conversations and no sufficient labels regarding the characteristics of the speakers.

A fair speaker diarization algorithm should process indifferently people from various minorities or sub-groups of the population. Taking that into account, in this paper we propose a new protocol designed to evaluate speaker diarization fairness. It is applied on single speaker utterances so the variability between our experiment is reduced to the factor studied. This protocol is, as far as we know, the first ever published to study speaker diarization fairness.

### 3.2. Protocol

Instead of considering conversations with a lot of internal properties that can impact diarization performances, a state-of-the-art diarization pipeline is applied on utterances spoken by only one speaker. This means the algorithm is expected to detect one or multiple segments spoken by only one identified speaker. If the system detects more than one speaker, it is an error.

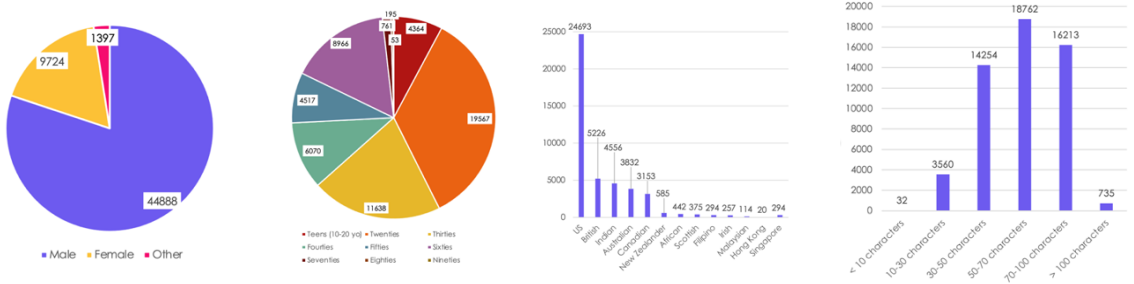

**Fig. 1.** Distribution of the different gender, age, accent and sentence length classes in the english CommonVoice dataset

For each utterance, the total number of speakers detected is reported. It can be 0 if no speaker has been tagged by the diarization.

We studied the impact of changing one of the following factors on speaker diarization results: gender, age, accent and spoken sentence length.

For each experiment performed on $N$ spoken utterances, let's define $U = \{u_i, i \in [\![1; N]\!]\}$ as the set of utterances and $X$ a random variable from $U$ to $\mathbb{N}$. $X$ corresponds to the number of speakers detected by our speaker diarization algorithm.

By applying this algorithm to all the $u_i$, the statistical distribution associated to $X$ is estimated and reported as follow:

- $p_0 = P(X = 0)$ corresponds to the case where no speaker has been detected and the diarization outputs no speech segment,
- $p_1 = P(X = 1)$ corresponds to the case where the accurate number of speakers has been identified by the diarization algorithm,
- $p_+ = P(X \geq 2)$ corresponds to the case where too many speakers have been identified by the diarization algorithm.

For each of these probability values, the standard error was computed and the 99% confidence margins with the following formula.

$$\varepsilon(p) = 2.58 \sqrt{\frac{p(1-p)}{N}} \quad (1)$$

Given this formula, we know that there is a 99% probability that the actual result $\bar{p}$ is as follow:

$$p - \varepsilon(p) < \bar{p} < p + \varepsilon(p) \quad (2)$$

### 3.3. Scoring Method

We propose to introduce the Diarization Fairness Rate (DFR) to compute speaker diarization fairness over a particular criterion as follow:

$$DFR = p_1 = 1 - p_0 - p_+ \quad (3)$$

This definition allows to separate the two terms $p_0$ and $p_+$, that respectively correspond to the bias due to the voice activity detection and segmentation for the first term and the bias due to the clustering in the second term.

### 3.4. Dataset and method used

We applied this protocol on the 9$^{th}$ version of the English CommonVoice dataset [12]. Indeed, this collaborative and open-source dataset contains more than 2000 hours of utterances spoken by a large variety of speakers.

Every sentence has been pronounced by a voluntary speaker who then self-reported his age, gender, accent, and the pronounced sentence.

We acknowledge the fact that this dataset, as it is collaboratively built, may have some flaws and implicit biases, but it is, as far as we know, the only dataset with the sufficient annotation labels to perform a fairness study on speaker diarization.

Although this dataset is not balanced on every label, it contains a very high number of utterances for each of the necessary classes to study the fairness of speaker diarization.

A speaker diarization algorithm, composed of a C-RNN voice activity detection [17], a x-vector extractor [18] trained on VoxCeleb 1 & 2 [19] and CnCeleb, and a Bayesian HMM clustering method [20] was applied on every utterances. All these components are achieving state-of-the-art results on their respective modules. They all have been open sourced by their authors, making this research completely reproducible.

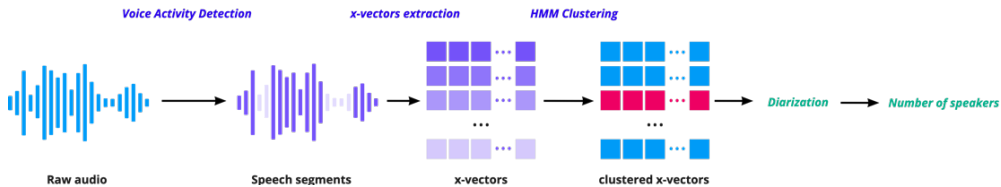

**Fig. 2.** Distribution of the different gender, age, accent and sentence length classes in the english CommonVoice dataset

# 4. RESULTS

**Table 1.** Obtained fairness results for the different experiments performed on gender, age, accent and sentence length

| | Gender | | | Sentence Length | | | | | | Age | | | | | | | | |
|---|---|---|---|---|---|---|---|---|---|---|---|---|---|---|---|---|---|---|
| | Male | Female | Other | < 10 char. | 10-30 char. | 30-50 char. | 50-70 char. | 70-100 char. | > 100 char. | Teens | Twenties | Thirties | Forties | Fifties | Sixties | Seventies | Eighties | Nineties |
| $p_0$ | 1.31 | 0,77 | 0,79 | 25,00 | 13,20 | 1,08 | 0,05 | 0,00 | 0,00 | 1,76 | 1,55 | 1,23 | 0,72 | 1,00 | 0,62 | 0,13 | 0,00 | 0,00 |
| $p_1$ | 88.20 | 93.39 | 94.27 | 62,50 | 77,39 | 87,99 | 90,60 | 91,32 | 92,24 | 91,93 | 87,45 | 89,53 | 91,17 | 88,19 | 90,92 | 90,28 | 96,41 | 77,36 |
| $p_+$ | 10.49 | 5.84 | 4.94 | 12,50 | 9,41 | 10,93 | 9,34 | 8,68 | 7,76 | 6,30 | 11,00 | 9,25 | 8,11 | 10,81 | 8,46 | 9,59 | 3,59 | 22,64 |

| | Accent | | | | | | | | | | | | |
|---|---|---|---|---|---|---|---|---|---|---|---|---|---|
| | US | British | Indian | Australian | Canadian | New-Zealander | African | Scottish | Filipino | Irish | Malaysian | Hong-Kong | Singapore |
| $p_0$ | 1,17 | 2,05 | 0,92 | 0,63 | 1,93 | 0,00 | 0,23 | 0,53 | 0,00 | 0,78 | 0,00 | 0,00 | 0,00 |
| $p_1$ | 90,98 | 89,13 | 84,64 | 88,75 | 91,69 | 88,55 | 83,48 | 87,73 | 92,24 | 87,16 | 93,86 | 90,00 | 68,37 |
| $p_+$ | 7,85 | 8,82 | 14,44 | 10,62 | 6,37 | 11,45 | 16,29 | 11,73 | 7,76 | 12,06 | 6,14 | 10,00 | 31,63 |

**Table 2.** Margins of error for the different experiments performed on gender, age, accent and sentence length

| | Gender | | | Sentence Length | | | | | | Age | | | | | | | | |
|---|---|---|---|---|---|---|---|---|---|---|---|---|---|---|---|---|---|---|
| | Male | Female | Other | < 10 char. | 10-30 char. | 30-50 char. | 50-70 char. | 70-100 char. | > 100 char. | Teens | Twenties | Thirties | Forties | Fifties | Sixties | Seventies | Eighties | Nineties |
| $\varepsilon(p_0)$ | 0,14 | 0,23 | 0,61 | 19,75 | 1,46 | 0,22 | 0,04 | - | - | 0,51 | 0,23 | 0,26 | 0,28 | 0,38 | 0,21 | 0,34 | - | - |
| $\varepsilon(p_1)$ | 0,39 | 0,65 | 1,60 | 22,08 | 1,81 | 0,70 | 0,55 | 0,57 | 2,55 | 1,06 | 0,61 | 0,73 | 0,94 | 1,24 | 0,78 | 2,77 | 3,44 | 14,83 |
| $\varepsilon(p_+)$ | 0,37 | 0,61 | 1,50 | 15,08 | 1,26 | 0,67 | 0,55 | 0,57 | 2,55 | 0,95 | 0,58 | 0,69 | 0,90 | 1,19 | 0,76 | 2,75 | 3,44 | 14,83 |

| | Accent | | | | | | | | | | | | |
|---|---|---|---|---|---|---|---|---|---|---|---|---|---|
| | US | British | Indian | Australian | Canadian | New-Zealander | African | Scottish | Filipino | Irish | Malaysian | Hong-Kong | Singapore |
| $\varepsilon(p_0)$ | 0,18 | 0,51 | 0,37 | 0,33 | 0,63 | - | 0,58 | 0,97 | - | 1,41 | - | - | - |
| $\varepsilon(p_1)$ | 0,47 | 1,11 | 1,38 | 1,32 | 1,27 | 3,40 | 4,56 | 4,37 | 4,03 | 5,38 | 5,80 | 17,31 | 7,00 |
| $\varepsilon(p_+)$ | 0,44 | 1,01 | 1,34 | 1,28 | 1,12 | 3,40 | 4,53 | 4,29 | 4,03 | 5,24 | 5,80 | 17,31 | 7,00 |

## 4.1. Gender bias

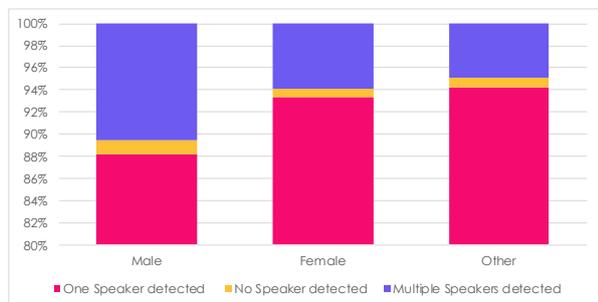

**Fig. 3.** Number of speakers detected according to the declared gender

Figure 3 presents the experiment concerning the declared gender of the speaker. The results show significant disparities between the scores obtained on male voices and the score obtained on female and other voices. 5% less male voices are correctly labelled by speaker diarization. About 5% more recordings were detected as spoken by multiple speakers among men than among women.

## 4.2. Age bias

This experiment on the age of the speaker demonstrates a relative fairness of our speaker diarization baseline since all age ranges seems to be equally treated. The small disparities between our results must be mitigated by the calculated error margins. For instance, for people in their nineties, speaker diarization seems to be less efficient but it is also the less represented class in the study. Therefore, the error margin of 14.83% prevent us from concluding anything but the probable fairness of speaker diarization over all age ranges.

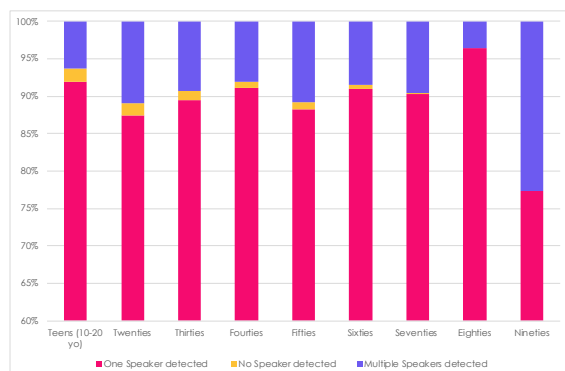

**Fig. 4.** Number of speakers detected according to the age

## 4.3. Accent bias

For this experiment, the spoken utterances pronounced by Welsh people were removed from the dataset as there were only three which is not sufficient to obtain reliable statistics.

The results on accent highlighted interesting disparities regarding the ability for speaker diarization to identify non-native English speaker as one single speaker. For speakers with African and Singaporean accents, the DFR is much lower. This echoes with the results presented in [10] about racial disparities in other voice applications.

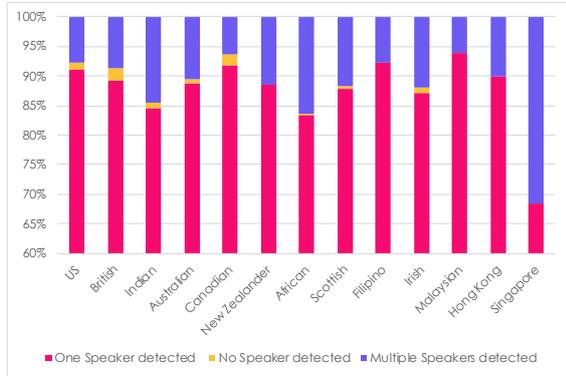

**Fig. 5.** Number of speakers detected according to the reported accent

### 4.4. Spoken Sentence Length bias

This last experiment on the impact of sentence length clearly shows that the shorter the spoken sentence, the higher the probability that it is not detected and therefore not assigned to any speaker. If this is not directly a bias in the sense that it can discriminate a certain category of people, this can question the language independency of speaker diarization. A language in which sentences tend to be shorter would produce more diarization errors. As it might be expected, this particular result is a direct consequence of the choice of the voice activity detection system [17].

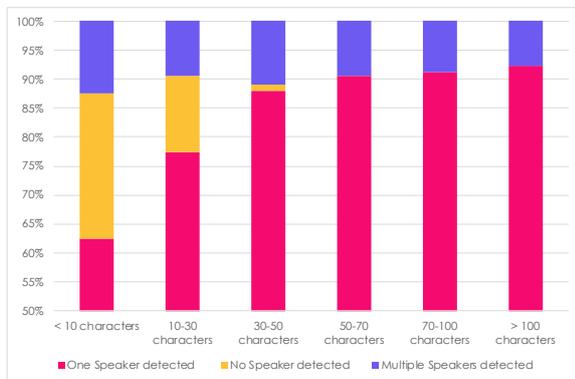

**Fig. 6.** Number of speakers detected according to the length of the utterance spoken

## 5. DISCUSSION

### 5.1. Results interpretation

The results show notable differences in the diarization performances depending on the gender, and accent of the person speaking.

The presented results and the DFR scores obtained appear as a good indicator for fairness evaluation, but they should be considered along with Diarization Error Rate (DER) scores. Indeed, in the proposed protocol, performing no diarization and always labelling one speaker would produce an ideal 100% DFR score.

### 5.2. Origin of the detected biases

It is hard to determine the origin of speaker diarization biases but here are some directions for further analysis.

First, one should consider the training set used for the x-vector embedding extractor. VoxCeleb and CnCeleb datasets contain interviews and speeches videos extracted from YouTube. These are relatively equilibrated datasets and there is no clear correlation between the distribution of the training set and the results presented in this paper. However, we don't have the labelled accents of the speakers used in our x-vectors training set.

According to our results, same speaker male voices are more likely to be clustered in a wrong multiple speakers' scenario than a female voice. One can explain this 4% difference by looking at the physiologic disparities between male and female pitch. It is well known that women have a broader frequency range than men [21]. This could be a possible explanation to why women are easier to cluster accurately during a single speaker diarization process.

### 5.3. Limitations of the proposed protocol

Although this protocol produced promising results about speaker diarization fairness, it is not yet perfect and has to be improved through future research.

This protocol was applied on utterances spoken by one speaker only. It is likely that the fairness results differ when looking at conversations performed by multiple speakers. That is why the proposed protocol should be extended to handle these cases.

The proposed protocol is also not method independent. Depending on the voice activity detection and clustering systems used, one can obtain different results.

## 6. CONCLUSION

In this article we proposed a new protocol to evaluate speaker diarization fairness and determine some of its biases. We also introduced a new metric, the DFR, and set some guidelines for future fairness studies.

We demonstrated the influence of the gender and the accent of the speaker as well as the length of the spoken sentence on diarization performances.

In the future, we plan to improve our protocol and study the fairness of audio-visual speaker diarization [14], [22] to determine whether the addition of visual information would contribute to reduce the observed biases.

## 7. ACKNOWLEDGMENTS

This research has received full financial support from the company Newsbridge (https://newsbridge.io/).